\documentclass[sigplan,10pt]{acmart}
\usepackage{soul}

\acmConference[PLDI'22]{ACM SIGPLAN Conference on Programming Language Design and Implementation}{June 13--17, 2022}{San Diego, CA, USA}
\acmYear{2022}
\acmISBN{} 
\acmDOI{} 
\startPage{1}

\setcopyright{none}

\usepackage{booktabs}   
\usepackage{subcaption} 

\begin{document}

\title[Abstract Interpretation on E-Graphs]{Abstract Interpretation on
E-Graphs}         


\author{Samuel Coward}
\affiliation{
  \department{Electrical and Electronic Engineering}              
  \institution{Imperial College London}            
}
\email{s.coward21@imperial.ac.uk}          

\author{George A.~Constantinides}
\affiliation{
  \department{Electrical and Electronic Engineering}              
  \institution{Imperial College London}            
}
\email{g.constantinides@imperial.ac.uk}         

\author{Theo Drane}
\affiliation{
  \institution{Intel Corporation}           
  \city{Folsom}
  \country{USA}                   
}
\email{theo.drane@intel.com}         


\maketitle

\section{Introduction} \label{sec:intro}

Recent e-graph applications have typically considered concrete semantics of expressions, where the notion of equivalence stems from concrete interpretation of expressions~\cite{Panchekha2015AutomaticallyExpressions,Wang2020SPORES:Algebra}. However, equivalences that hold over one interpretation may not hold in an alternative interpretation. Such an observation can be exploited. We consider the application of abstract interpretation to e-graphs, and show that within an e-graph, the lattice meet operation associated with the abstract domain has a natural interpretation for an e-class, leading to improved precision in over-approximation. In this extended abstract, we use Interval Arithmetic (IA) \cite{Neumaier1991IntervalEquations,Jaulin2002AppliedAnalysis} to illustrate this point.

IA is commonly used to provide tight bounds on expressions or numerical program outputs. This is useful across numerical hardware/software design and verification, providing guarantees that exceptional behaviour is never encountered and enabling deeper optimizations. 
In IA, every numerical expression is associated with an (real or floating-point) interval rather than a single numerical value and for each operation $f$ in the arithmetic, the natural interval extension operation is associated, where we assume that all infimums and supremums exist:
\begin{equation}
\label{extension}
f(X_1,..,X_k) = \left [\inf_{x_i \in X_i} f(x_1,...,x_k), 
\sup_{x_i \in X_i} f(x_1,...,x_k) \right]
\end{equation}
Let $\llbracket e \rrbracket$ denote the natural interval extension of an expression, produced by structural induction on the expression syntax.

A key limitation of IA is the dependency problem, as illustrated through the following example.
For $x\in [0,1]$ a standard interpretation gives:
\begin{equation} \label{eqn:ia_limits}
\llbracket x-x \rrbracket = [0,1]-[0,1] = [0-1, 1-0] = [-1,1].   
\end{equation}
 Implementing IA using e-graphs, helps to mitigate the effect of the dependency problem as we shall see in \S \ref{sec:approach}.

Among existing tools using IA to obtain tight bounds \cite{Martin-Dorel2016ProvingCoq, Daumas2005GuaranteedArithmetic}, Gappa \cite{Daumas2010CertificationOperators}, a tool for fixed and floating-point error analysis, is particularly relevant as it deploys a set of term rewrites in order to tackle the dependency problem.
 

\section{Approach}\label{sec:approach}
We implement IA for real arithmetic expressions on top of the extensible \texttt{egg} library \cite{Willsey2021Egg:Saturation} as an e-class analysis. Following a standard approach, we represent real intervals by pairs of floating-point values, conservatively approximating real operations by rounding away from zero, a technique known as `outwardly rounded IA' \cite{Moore2009IntroductionAnalysis,Kulisch1981ComputerPractice}, however our examples in this abstract are presented as real numbers for simplicity. 

The key insight of our work is that expressions that are equivalent in the concrete interpretation, and hence can belong to the same e-class in an e-graph, may differ in their abstract interpretation. Despite this difference in abstract interpretation, the soundness of the two or more different abstract interpretations of concrete-equivalent expressions, implies that they may be combined via the meet operation associated with the abstract lattice~\cite{Cousot1977AbstractFixpoints}, producing a more precise approximation.

A trivial example of this process derives from Eqn.~\ref{eqn:ia_limits}. Consider the expression $\texttt{x - x}$, together with the rewrite rule $\texttt{x - x} \to \texttt{0}$. This implies that $\texttt{x-x} \cong \texttt{0}$, where $\cong$ denotes concrete-equivalence. In an e-graph, an e-class for this expression would contain two nodes, corresponding to the equivalent expressions. The interval interpretation of these two expressions is $[-1,1]$ and $[0,0]$, respectively, and as a result we may conclude that {\em both} expressions lie in $[-1,1] \cap [0,0]$, where intersection is the meet operation of the interval lattice.

To describe interval propagation throughout the e-graph, define $C$ to be the set of e-classes, and $\mathcal{N}_c$ the set of e-nodes contained in $c\in C$. With each e-class, we associate a pair of floating-point values $(\underline{X}, \overline{X})$ to represent a real interval, which we denote $\llbracket c \rrbracket = [\underline{X}, \overline{X}]$.


Similarly interpret a $k$-arity e-node $n$ of function $f$ with children classes $c_{n,1}, ..., c_{n,k}$, as:
\begin{equation}
\label{interpnode}
\llbracket n \rrbracket = f \left (\llbracket c_{n,1} \rrbracket, ..., \llbracket c_{n,k} \rrbracket\right)    
\end{equation}
via the natural interval extension of the function of the e-node $n$, as per Eqn.~\ref{extension}. 0-arity e-nodes represent constants, associated with degenerate intervals containing a single value, or variables, accompanied by user specified intervals. 

For acyclic e-graphs, it is trivial to propagate the known intervals upwards through the e-graph using Eqn.~\ref{interpnode} together with the following novel tightening relationship, where meet $\sqcap$ is intersection for intervals.
\begin{equation}
\label{meettighten}
    \llbracket c \rrbracket = \bigsqcap_{n\in \mathcal{N}_c} \llbracket n \rrbracket
\end{equation}

Although described above in static terms for simplicity of exposition, it's important to note that Eqn.~\ref{meettighten} can be evaluated on the fly, as we discover further equivalences and grow the e-graph; these always make the approximation more precise, due to the monotonicity of the meet operation: just as the e-graph grows monotonically during construction, the associated abstract values within the e-graph will monotonically narrow, corresponding to more precise expression bounds. We also note that this property allows for computation with cyclic e-graphs, under which Eqns \ref{interpnode} and \ref{meettighten} form a fixpoint specification.

\begin{figure}
    \centering
    \subfloat[Initial e-graph: $\frac{y}{1+y}$] {\includegraphics[scale=0.41]{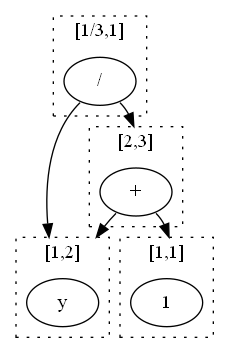}}
    \quad
    \subfloat[Applying $\frac{y}{1+y} \rightarrow \frac{1}{1+\frac{1}{y}}$] {\includegraphics[scale=0.41]{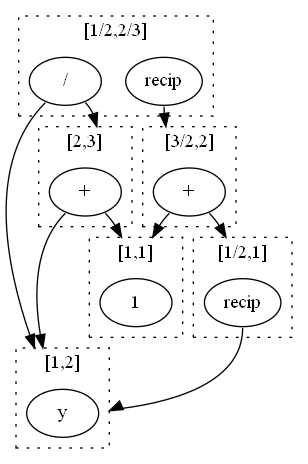}}
    \caption{Including intervals in the e-graph and rewriting to obtain tight bounds. \texttt{recip} is the reciprocal function.}
    \label{fig: e-graph_example}
\end{figure}

Figure \ref{fig: e-graph_example} demonstrates a non-trivial dependency problem example resolved by e-graph IA \cite{Moore2009IntroductionAnalysis}. The over-approximation in the expression, $\frac{y}{1+y}$, stems from the multiple occurrences of $y$ in the expression. By rewriting, a concrete-equivalent expression is discovered, in which $y$ only appears once, removing the dependency issue.

\section{Interval Implementation}

For this work we use a set of 23 rewrites. The basic arithmetic rewrites are commutativity, associativity, distributivity, cancellation and idempotent operation reduction across addition, subtraction, multiplication and division. The set also contains conditional polynomial rewrites for factorising using the quadratic formula and completing the square.
Lastly, we include two operator specific rewrites.
\begin{equation*}
    \sqrt{\texttt{a}} \texttt{-}\sqrt{\texttt{b}} \to \frac{\texttt{a-b}}{\sqrt{\texttt{a}} \texttt{+} \sqrt{\texttt{b}}}\hspace{1em},\hspace{1em}
    \frac{\texttt{a}}{\texttt{b}}         \to \frac{\texttt{b - (b-a)}}{\texttt{b}}
\end{equation*}

Using these rewrites we are able to tighten expression bounds on the simple expressions presented in Table \ref{tab:basic_results}. On these small examples \texttt{egg} runs in seconds, since interval calculations are relatively cheap. 

\begin{table}
    \centering
    \begin{tabular}{|c|r|r|c|}
         \hline
         Expression & Initial &  \multicolumn{1}{c|}{Improved} & Width Change\\
         \hline
         $x^2 -2x+1$             & [-2,3]   & [0,1]     & -80\%\\[7pt]
         $x(2-xy) - \frac{1}{y}$ & [-5,1.5] & [-4.5, 0] & -31\%\\[7pt]
         $\sqrt{x+1} - \sqrt{x}$ & $[0,\sqrt{2}]$ & $\left[\frac{1}{\sqrt{2}+\sqrt{3}}, \frac{1}{1+\sqrt{2}}\right]$ & -93\%\\[7pt]
         
         $\frac{x}{x+y}$         & $\left[\frac{1}{4}, 1\right]$ & $\left[\frac{1}{4}, \frac{3}{4}\right]$ & -33\% \\[7pt]
         \hline
    \end{tabular}
    \caption{Interval tightening via e-graphs, for $x,y \in [1,2]$. Width change=(improved width - initial width)/initial width.}
    \label{tab:basic_results}
\end{table}


To demonstrate an advantage of this approach, consider the following example, for $x\in [0,1], y\in [1,2]$, where the following concrete-equivalences are discovered via rewriting:
\begin{align}
                & 1 - \frac{2y}{x+y} &\in \left[-3,\frac{1}{3}\right] \label{eqn:rw_a}  \\
    &\cong  \frac{x-y}{x+y}    &\in [-2,0] \label{eqn:rw_b}\\
    &\cong  \frac{2x}{x+y} - 1 &\in [-1,1] \label{eqn:rw_c}.
\end{align}
The interval associated with the e-class containing these three expressions is $[-3,\frac{1}{3}] \cap [-2,0] \cap [-1,1] = [-1,0]$.
We observe that there is no need to find a single expression providing both bounds, which may in general be impossible. In more general abstract interpretations, we may find a set of expressions, each providing valuable and distinct information. 



\section{Conclusion and Future Work}\label{sec:results}
We presented the application of abstract interpretations to e-graphs, which has two key advantages. Due to constructive rewrite application deciding which rewrites to apply and in which order is not a concern in the e-graph, useful if the route to tightly bounding expressions is non-obvious.
There is also no constraint on the number of expressions that can provide relevant information in a given interpretation.

Further work will explore more complex problems, along with comparisons against existing tools such as Gappa \cite{Daumas2010CertificationOperators}, and results on relational domains. We will apply rewrite rule inference~\cite{Nandi2021RewriteSaturation} to explore the space of bound tightening rewrites. It may be further possible to exploit cyclic e-graphs on abstract domains by interpreting the fixpoint equations as defining an iterative numerical method such as the Krawczyk method~\cite{Moore2009IntroductionAnalysis} which may then be extracted from the e-graph.
Incorporating the technique into tools, such as Herbie~\cite{Panchekha2015AutomaticallyExpressions}, where bounds can be exploited, would demonstrate its value.

\bibliography{references}

\end{document}